\input harvmac
\input epsf

\newcount\figno
\figno=0
\def\fig#1#2#3{
\par\begingroup\parindent=0pt\leftskip=1cm\rightskip=1cm\parindent=0pt
\baselineskip=11pt
\global\advance\figno by 1
\midinsert
\epsfxsize=#3
\centerline{\epsfbox{#2}}
\vskip 12pt
{\bf Fig. \the\figno:} #1\par
\endinsert\endgroup\par
}
\def\figlabel#1{\xdef#1{\the\figno}}
\def\encadremath#1{\vbox{\hrule\hbox{\vrule\kern8pt\vbox{\kern8pt
\hbox{$\displaystyle #1$}\kern8pt}
\kern8pt\vrule}\hrule}}

\overfullrule=0pt

\noblackbox
\parskip=1.5mm

\def\Title#1#2{\rightline{#1}\ifx\answ\bigans\nopagenumbers\pageno0
\vskip0.5in
\else\pageno1\vskip.5in\fi \centerline{\titlefont #2}\vskip .3in}


\noblackbox
\parskip=1.5mm

  
\def\npb#1#2#3{{\it Nucl. Phys.} {\bf B#1} (#2) #3 }
\def\plb#1#2#3{{\it Phys. Lett.} {\bf B#1} (#2) #3 }
\def\prd#1#2#3{{\it Phys. Rev. } {\bf D#1} (#2) #3 }

\def\bb#1{{\tt hep-th/#1}}
\def\grqc#1{{\tt gr-qc/#1}}

\def\app#1#2#3{{\it Astropart. Phys. } {\bf #1} (#2) #3 }
\def\cqg#1#2#3{{\it Class. Quant. Grav. } {\bf #1} (#2) #3 }
\def\apj#1#2#3{{\it Astrophys. J.  } {\bf #1} (#2) #3 }



\def\dj{\hbox{d\kern-0.347em \vrule width 0.3em height 1.252ex depth
-1.21ex \kern 0.051em}}


\lref\rbar{J.D. Barrow and K. Kunze, {\it Inhomogeneous String Cosmologies,}
(\bb{9701085}.)}
\lref\rus{A. Feinstein, R. Lazkoz and M.A. V\'azquez-Mozo, \prd{56}{1997}{5166}
(\bb{9704173}).}
\lref\rclt{D. Clancy, J.E. Lidsey and R. Tavakol, {\it Effects of anisotropy and
spatial curvature in the pre-big bang scenario}, to appear in Phys. Rev. D (\grqc{9802052}).}
\lref\rtw{M.S. Turner and E.J. Weinberg, \prd{56}{1997}{4604} (\bb{9705035}).}
\lref\rbd{J.D. Barrow and M.P. Dabrowski, \prd{55}{1997}{630} (\bb{9608136}).}
\lref\rsfd{D. Clancy, J.E. Lidsey and R. Tavakol, \cqg{15}{1998}{257} 
(\grqc{9802053}).}
\lref\rqdlo{X. de la Ossa and F. Quevedo, \npb{403}{1993}{377} (\bb{9210021}).}
\lref\rch{C.B. Collins and S.W. Hawking, \apj{180}{1973}{317.}}
\lref\rn{D. Clancy, J.E. Lidsey and R. Tavakol, {\it Initial conditions in string
cosmology} (\grqc{9806065}).}
\lref\rbguv{A. Buonanno, K.A. Meissner, C. Ungarelli and G. Veneziano, \prd{57}{1998}{2543}
(\bb{9706221}).}
\lref\rwitt{E. Witten, \npb{443}{1995}{85} (\bb{9503124}).}
\lref\rpbb{G. Veneziano, \plb{265}{1991}{287\semi}
M. Gasperini and G. Veneziano, \app{1}{1993}{317} (\bb{9211021})\semi
G. Veneziano, {\it A simple/short introduction to Pre-big Bang physics/cosmology},
Preprint CERN-TH/98-43 (\bb{9802057}).} 
\lref\rgep{M. Gasperini and M. Giovannini, \plb{287}{1992}{56\semi}
R. Brustein and G. Veneziano, \plb{329}{1994}{429} (\bb{9403060})\semi
M. Gasperini, M. Maggiore and G. Veneziano, \npb{494}{1997}{315} (\bb{9611039})\semi
R. Brustein and R. Madden, \prd{57}{1998}{712} (\bb{9708046}).}
\lref\rven{G. Veneziano, \plb{406}{1997}{297} (\bb{9703150}).}
\lref\res{E.J. Copeland, A. Lahiri and D. Wands, \prd{50}{1994}{4868} (\bb{9406216})\semi
N.A. Batakis, A.A. Kehagias, \npb{449}{1995}{248} (\bb{9502007})\semi
J.D. Barrow and K. Kunze, \prd{56}{1997}{741} (\bb{9701085}).}


\line{\hfill EHU-FT/9806}  
\line{\hfill {\tt hep-th/9806167}}
\vskip 1.5cm

\Title{\vbox{\baselineskip 12pt\hbox{}
 }}
{\vbox {\centerline{Can the Type-IIB axion prevent Pre-big Bang inflation?}
}}

\vskip 0.8cm

\centerline{$\quad$ {A. Feinstein\foot{wtpfexxa@lg.ehu.es} 
and M.A. V\'azquez-Mozo\foot{wtbvamom@lg.ehu.es}
 }}

\medskip

\centerline{{\sl Dpto. de F\'{\i}sica Te\'orica}}
\centerline{{\sl Universidad del Pa\'{\i}s Vasco}}
\centerline{{\sl Apdo. 644, E-48080 Bilbao, Spain}}

 \vskip 1.8cm

\noindent
We look at the possibility of superinflationary behavior in a class of 
anisotropic Type-IIB superstring cosmologies in the context of Pre-big Bang 
scenario and 
find that there exists a rather narrow range of parameters for which these
models inflate. We then show that, although in general this behavior is left 
untouched by  the introduction of a Ramond-Ramond axion field through a 
$SL(2,{\bf R})$ rotation, there exists a particular class of axions 
for which inflation disappears completely. Asymptotic past initial 
conditions are briefly discussed, and some speculations on the possible 
extension of Pre-big Bang ideas to gravitational collapse are presented.


\Date{06/98}


The Pre-big Bang scenario developed recently within the context of string theory
represents an interesting alternative to the standard inflationary paradigm 
\refs\rpbb. In this
approach a massless scalar field, the dilaton, drives an accelerated expansion towards
what we normally think of as the initial cosmological singularity. It is believed that near the 
singularity
string non-perturbative physics takes over and ensures a smooth transiton to the 
next phase where the standard Friedman-Robertson-Walker expansion would take place. While 
until very recently 
most investigations were concentrated on the details of this transition, the 
so-called graceful exit problem \refs\rgep,
it has been lately realized that the problem of the susceptibility of the Pre-big Bang phase
to initial conditions remains the cornerstone of the scenario. 
Intuitively, the Pre-big Bang phase may be robust to small irregularities 
induced by
anisotropies and inhomogeneities. This belief is based on the fact that during the Pre-big
Bang phase the cosmological model undergoes a period of accelerated expansion so that 
irregularities, if any, are likely to be "inflated away". In this spirit, 
Veneziano and collaborators \refs\rven\rbguv\
have shown that sufficiently smooth regions of spacetime undergo inflation during the
Pre-big Bang phase in spite of the presence
of small deviations from homogeneity and isotropy. On the other hand, Turner and Weinberg
\refs\rtw\ have looked at the effects of the curvature on the
Pre-big Bang inflation and have found that large spatial curvature may postpone the onset
of inflation, interpreting this result as a restriction on the possible initial conditions.
To further study the effects of anisotropy and inhomogeneity
a number of exact anisotropic and inhomogeneous solutions have been found \refs\res\refs\rbd\
and general algorithms to construct these solutions were designed \refs\rus. 

In a recent paper Clancy {\it et al.} \refs\rclt\ considered a class of exact anisotropic
cosmological solutions which contains Bianchi types III, V and VI$_{h}$ models
focusing their study on the combined effects of spatial anisotropy and curvature on the
Pre-big Bang period. They have found that large regions in the parameter space exist
where the occurence of inflation is prevented, and that the conditions for successful
inflation, as compared with the negatively curved FRW model, are stringent.

Now, although the Pre-big Bang scenario is string theory induced, surprisingly, most of the
work in Pre-big Bang cosmology is done in the so-called Brans-Dicke sector,
where the only  addition to general relativity is the presence of a dilaton field.
Yet, the low energy sector of the string theory may generally include various massless
fields, and for example in the case of the Type-IIB superstring one expects to find along
with the scalar dilaton a pseudoscalar axion field $\chi$ (among other fields) in the
Ramond-Ramond sector of the theory. This pseudoscalar field is rather interesting due to its
peculiar coupling to the dilaton and its presence, in principle, might play a crucial role 
in damping Pre-big Bang inflation. 

The main purpose of this letter is to study the possible effects derived from the
presence of the pseudoscalar axion field in the Type-IIB superstring spectrum 
on the occurence of the Pre-big Bang inflation in
anisotropic cosmological models. To do so, we consider an exact class of anisotropic
Kantowski-Sachs cosmologies; first, in the presence of the dilaton field only, and then
adding the pseudoscalar axion field into the picture. In the case  of the
dilaton field alone, we confirm the behaviour found in \refs\rclt\ in that the 
range of
the parameters producing inflationary behaviour is quite narrow. The introduction of the
pseudoscalar axion field, however, may produce a devastating effect on inflation reducing
the range of the possible parameters to an empty set. Fortunately for the scenario, the
effect is restricted to a special class of axions. Our test cosmologies,
Kantowski-Sachs models, are of interest in this context for they combine the effects of
anisotropy and non-trivial curvature, furnishing the conclusions we arrive at with 
quite a general character. On the other hand, these models are simple enough so that 
one can integrate the exact solutions explicitly and study their behavior 
analytically.

We start with the following family of Kantowski-Sachs metrics in string frame
\eqn\family{
ds^2=-\left({2\eta\over t}-k\right)^{-\alpha}dt^2 + 
\left({2\eta\over t}-k\right)^{\beta} dx^2 + \left({2\eta\over t}-k\right)^{1-\alpha}
t^2[d\theta^{2}+f_{k}(\theta)^2 d\varphi^{2}]
}
coupled to the dilaton field
\eqn\dilaton{
\phi={1\over 4}(\beta-\alpha)\log\left({2\eta\over t}-k\right).
}
This is an exact solution of dilaton gravity provided the constants $\alpha$ and 
$\beta$ satisfy 
$$
\alpha^2+\beta^2=2.
$$ 
Here $k$ is the spatial curvature ($k=-1,0,1$) and 
$$
f_{k}(\theta)=\left\{\matrix{\sin{\theta} & k=1 \cr \theta & k=0 \cr
\sinh{\theta} & k=-1 } \right. .
$$
The family of metrics defined by eqs. \family\ and \dilaton\ can be readily obtained by 
starting with vacuum 
Kantowski-Sachs cosmologies (corresponding here to $\alpha=\beta=1$)
and dressing those with a massless scalar field with the subsequent frame change along the 
first steps of a general algorithm given in \refs\rus.

Incidentally, the solution \family\ is equivalent to the one recently given by Barrow and
Dabrowski \refs\rbd, who used different parametrization and integrated the equations of motion
directly. The explicit simple solution integrated by these authors (eqs. 3.9-3.12 of
ref. \refs\rbd) are obtained just by setting $\alpha=0$ in eq. \family\ and taking into 
account the different normalization of the dilaton fields.

In ref. \refs\rsfd\ the scale factor
duality of anisotropic Kantowski-Sachs models was studied with the result that 
the only discrete symmetry of the dynamical equations consisted in inverting the
scale factor in the $x$-direction. In \family\ and \dilaton\ this corresponds 
to flipping the sign of $\beta$ ($\beta\rightarrow -\beta$).

At this point we feel that it is important to clarify some differences  between 
scale factor duality and T-duality. In scale factor duality 
the inversion of one or more scale factors in the metric leaves the equations of
motion invariant (or equivalently modifies the low-energy action by a total
derivative). T-duality, on the other hand, is a transformation of the metric, not
necessarily only of the scale factors, leading to a new solution of the low-energy 
field equations which corresponds to an equivalent string theory. 
In the metric \family\ we have four Killing vectors
$$
\eqalign{\xi_{1}&=\sin{\varphi}\partial_{\theta}+{f'_{k}(\theta)\over f_{k}(\theta)}
\cos{\varphi}\partial_{\varphi} \cr
\xi_{2}&=\cos{\varphi}\partial_{\theta}-{f'_{k}(\theta)\over f_{k}(\theta)}
\sin{\varphi}\partial_{\varphi} \cr
\xi_{3}&=\partial_{\varphi} \cr
\xi_{4}&=\partial_{x}.
}
$$
It is straightforward to see that the Killing vector $\xi_{4}$ commutes with all the 
others. Compactifying the $x$ direction we have, in the particular case of 
the positively curved Kantowski-Sachs 
model ($k=1$), that the total isometry group is $U(1)_{x}\times
SO(3)$, and its maximal abelian subalgebra is generated by $\partial_{x}$ and 
$\partial_{\varphi}$.
T-dualizing with respect to this pair of commuting Killing vectors we arrive at a dual
manifold in which the only Killing vectors left are $\partial_{\tilde{x}}$ and 
$\partial_{\tilde{\varphi}}$, with $\tilde{x}$ and $\tilde{\varphi}$ being the dual 
coordinates. Any other isometry not
commuting with $\xi_{3}$ and $\xi_{4}$ will be lost in the sense that it will not be locally 
realized in the dual
manifold any longer \refs\rqdlo. Therefore, the degree of homogeneity may be reduced by 
just performing a 
T-duality transformation \refs\rus. Consequently, for the model described by the line
element \family\ with a cyclic $x$ coordinate, the only T-duality 
transformation leaving the spatial symmetry group intact is the one performed along the 
compactified isometric direction $x$, which in this case is formally identical to the scale 
factor duality of the metric \family. In the case of scale factor duality, however, 
there is no 
need to impose the compactness of the $x$-coordinate since the ``dual'' model does 
not necessarily have to be equivalent to the original one as complete string theories.

We now look at the possible range of the parameters leading to inflationary periods for the 
metric \family. Due to the invariance of the low-energy field equations under time reversal,
and in order to keep expressions simple, we will use positive time to label the Pre-big Bang
phase, so that the approach to the Big-bang singularity from this phase corresponds to  
$t\rightarrow 0^{+}$ and the asymptotic past infinity is mapped into the asymptotic future
$t\rightarrow \infty$. By expanding the solution near $t=0$ and using co-moving time $T$ we 
find the following form for the metric
$$
ds^2=-dT^2+a^{2}_{1} T^{-{2\beta\over \alpha+2}}dx^{2}+a^{2}_{2} T^{2{\alpha+1\over \alpha+2}}
[d\theta^2+f_{k}(\theta)^2 d\varphi^2]
$$
with $a_{1}^2$, $a_{2}^2$ two constants and the dilaton being
$$
\phi(T)=-{1\over 2}{\beta-\alpha\over \alpha+2}\log T.
$$

Close to $T=0$ we find that the scalar curvature scales
with time as $R\approx T^{-2}$, while the behavior of the dilaton does depend on 
the particular values of $\alpha$ and $\beta$ and the effective string coupling in this
limit is
$$
g_{\rm eff}=e^{\phi(T)}\approx T^{-{1\over 2}{\beta-\alpha\over \alpha+2}}.
$$
When $\alpha>\beta$, $g_{\rm eff}\rightarrow 0$ and we have that
near $T=0$ the string theory is weakly coupled in spite of having a curvature singularity.
In the other case $\alpha<\beta$ we have the reverse situation in which both the 
curvature and the effective string coupling blow up when the limit $T=0$ is 
approached. The degenerate
case $\alpha=\beta=\pm 1$ corresponds to have a vanishing dilaton.
Since one of the characteristics of the Pre-big Bang scenario is that the approach to the 
curvature singularity from the Pre-big Bang phase is described by a strongly coupled 
string theory, we will restrict in the following our attention to the latter case 
$\beta>\alpha$.
\fig{The thick line represents the set of parameters for which inflation 
is possible.}{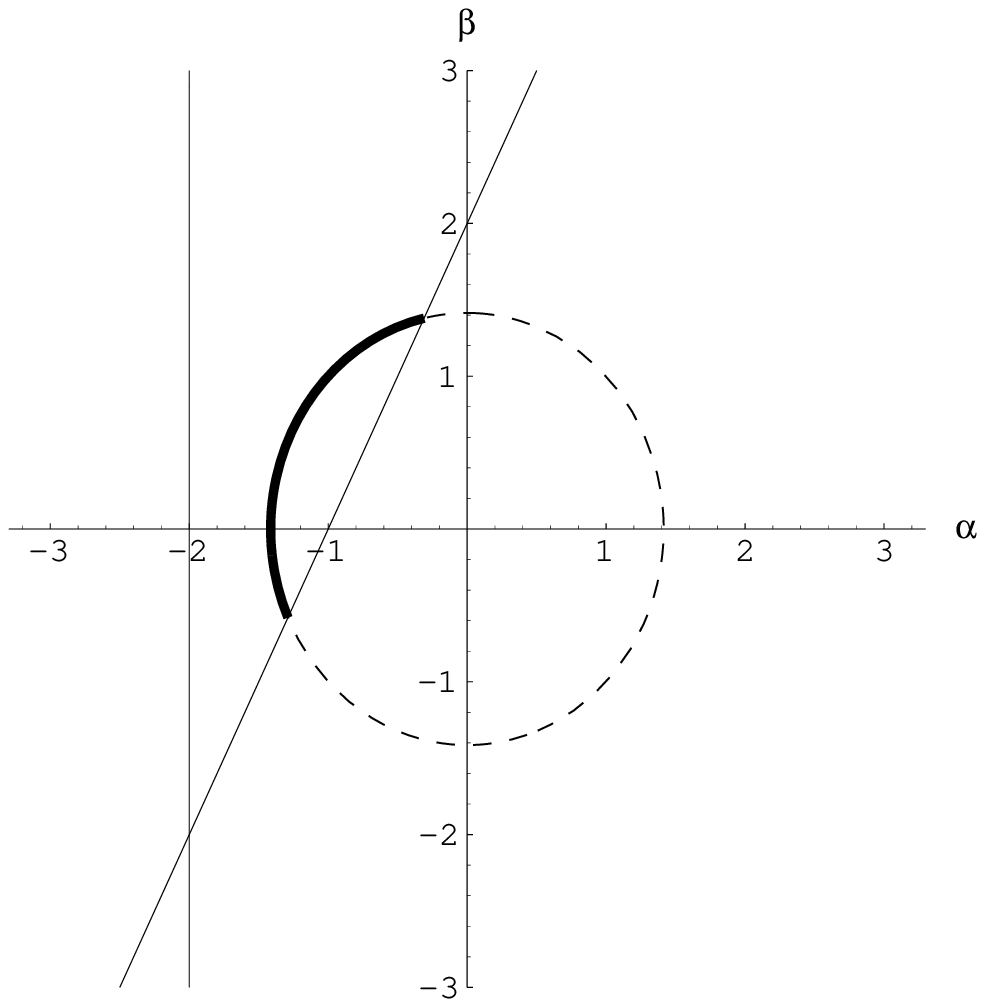}{2truein}

Let us now define the averaged scale factor
$$
\tilde{R}(T)=[a_{1}a_{2}^2]^{1\over 3}T^{2\alpha-\beta+2\over 3(\alpha+2)}
$$
Pre-big Bang cosmology is characterized by the superinflationary behavior achieved in
the case of negative exponents of the scale factor. Thus, the conditions for ``averaged''
inflation are
$$
{2\alpha-\beta+2\over 3(\alpha+2)}<0, \hskip 1cm \alpha^2+\beta^2=2
$$
In fig. 1 we represent the points in parameter space satisfying both conditions. 
For all points on the circumference lying above the line $2\alpha-\beta+2=0$
the exponent of the co-moving time in the averaged scale factor will be negative 
and thus will represent metrics undergoing accelerated expansion as $T$ approaches zero.
On the other hand in fig. 2 we plot those points for which {\it both} scale
factors undergo inflation in the $T\rightarrow 0$ limit. 
\fig{Set of points for which {\it both} scale factors inflate (thick line).}{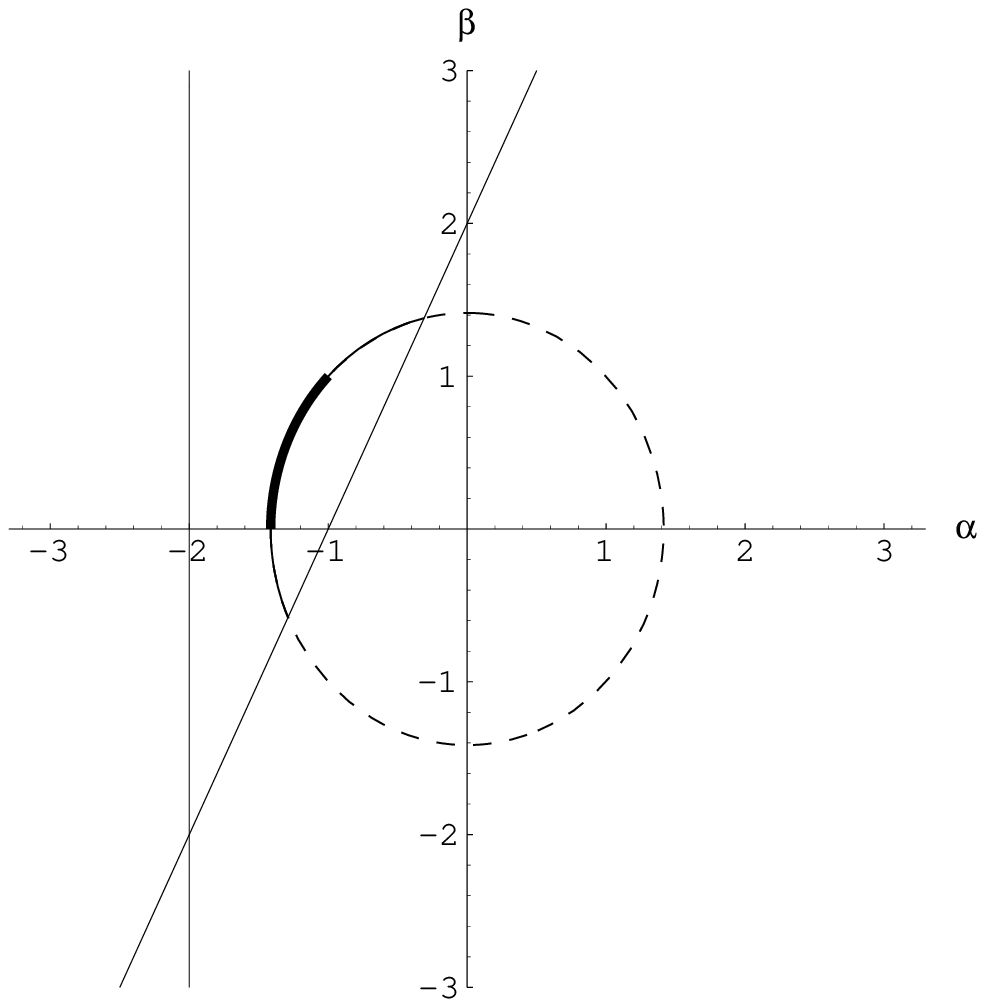}{2truein}

To introduce the Type-IIB axion into the picture we perform a $SL(2,{\bf R})$ rotation on 
the graviton-dilaton system
\family, \dilaton. From \refs\rus\ we find the new metric, dilaton and axion to be
\eqn\rot{
\eqalign{ds'^{\,2}&=[d^2+c^2 e^{-2\phi(t)}]^{1/2}\,ds^2 \cr
e^{-\phi'(t)}&={e^{-\phi(t)}\over [d^2+c^2 e^{-2\phi(t)}]^{1/4}} \cr
\chi'(t)&={bd+ac\, e^{-2\phi(t)}\over [d^2+c^2 e^{-2\phi(t)}]^{1/4}},
}}
where $a,b,c,d\in{\bf R}$ and $ad-bc=1$. 
Since as the result of the rotation \rot\ the metric changes conformally, it can 
affect in principle the inflationary behavior near the singularity. Notice, however, that
four-dimensional $SL(2,R)$ transformations do not invert the string coupling 
constant, as it is the case in ten-dimensions (cf. \refs\rwitt).

Let us first analyze the family of transformations with $d=0$, for which the metric 
transforms non-trivially near the singularity. The conformal transformation
of the metric is just given by 
$$
ds'^{\,2}_{d=0}=|c| e^{-\phi(t)} ds^2.
$$
Passing to co-moving time and expanding for small $T$ we have
$$
ds'^{\, 2}_{d=0}=-dT^2+b^{2}_{1}T^{-2{\alpha+3\beta\over 3\alpha+\beta+8}}dx^{2}+
b^{2}_{2}T^{2{3\alpha+\beta+4\over 3\alpha+\beta+8}}
[d\theta^2+f_{k}(\theta)^2 d\varphi^2]
$$
where we have absorbed the $c$-dependence by redefining 
the constants $a_{1}^2$ and $a_{2}^2$. Evaluating the average scale factor we find
\eqn\afd{
\tilde{R}'(T)_{d=0}=
[b_1 b^{2}_{2}]^{1\over 3}T^{8+5\alpha-\beta\over 3(8+3\alpha+\beta)}
}
for all $\alpha$, $\beta$ such that $\alpha^2+\beta^2=2$. 

To have inflation the exponent of 
the averaged scale factor $\tilde{R}'(T)$ in \afd\ 
must be negative. Plotting these conditions in fig. 3 we find that 
there are no allowed values of the parameters $\alpha$ and $\beta$ for which the 
model undergoes inflation as $T\rightarrow 0$. 
We thus arrive at the conclusion that the introduction of the Type-IIB axion via a 
$SL(2,{\bf R})$ rotation of the graviton-dilaton system with $d=0$ completely topples 
the Pre-big Bang inflation in the family \family\ of Kantowski-Sachs metrics.
\fig{$SL(2,{\bf R})$-induced dilatons with $d=0$ remove averaged 
inflation.}{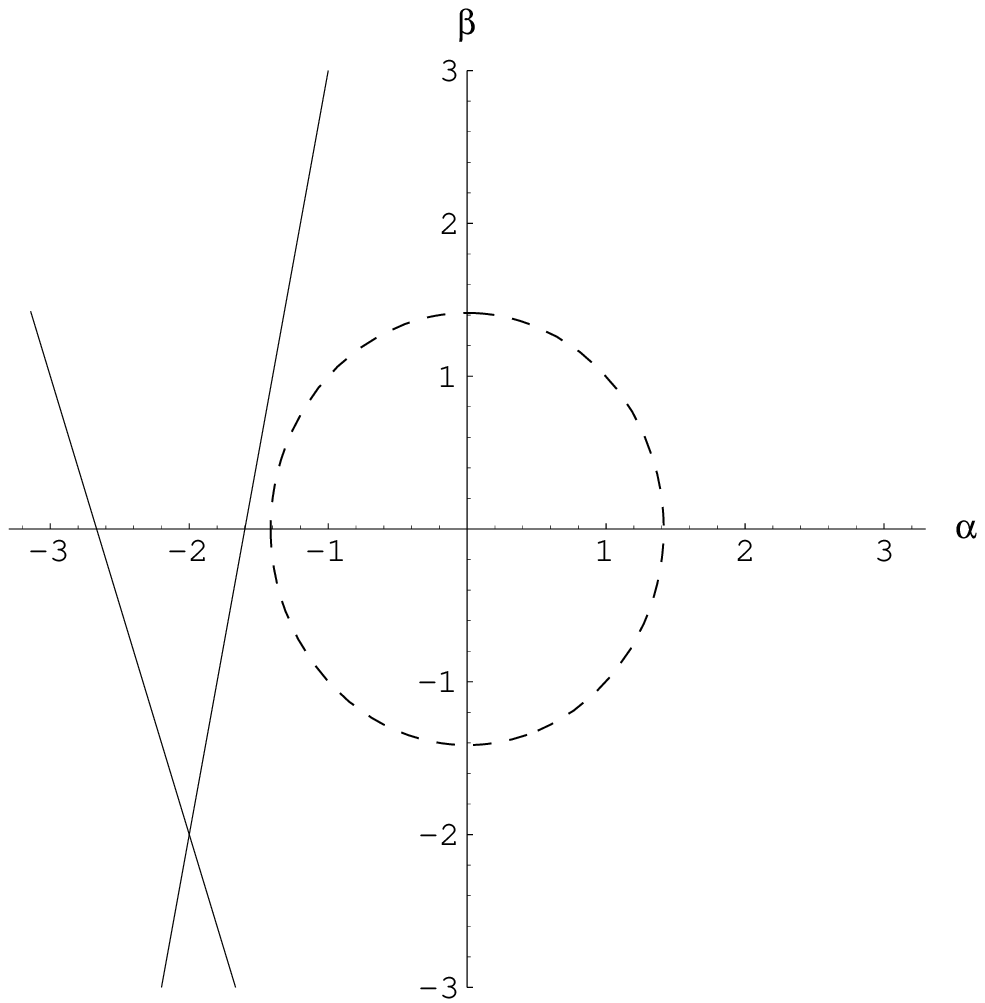}{2truein}

Fortunately, things are different when we introduce the axion using a 
$SL(2,{\bf R})$ with $d\neq 0$. The $T\rightarrow 0$ behavior of the conformal 
factor multiplying the metric will depend on the limit of the dilaton field $\phi(T)$ close 
to the singularity. Since we are restricting ourselves to the case $\alpha<\beta$, the dilaton
field blows up as $T=0$ is approached, and the effect of the $SL(2,{\bf R})$
rotation on the metric is just equivalent in this limit to a constant rescaling
of the metric tensor. Reabsorbing constant factors we arrive at a metric equivalent
to the one we started with
$$
ds'^{\,2}_{d\neq 0}=-dT^2+b'^{2}_{1} T^{-{2\beta\over \alpha+2}}dx^{2}+
b'^{2}_{2} T^{2{\alpha+1\over \alpha+2}}[d\theta^2+f_{k}(\theta)^2 d\varphi^2],
$$
where, as above, $\alpha^2+\beta^2=2$ and $\alpha<\beta$. Thus, generically, the 
conditions for inflation are left unchanged by the presence of a non-trivial pseudoscalar 
axion field. We will not dwell here on the influence of the axion field on the onset of
inflation \refs\rtw, but it is clear that even when $d\neq 0$ the axion might play a role
in determining the period of time in which inflation may ocurr away from $T\rightarrow 0^+$.

Another interesting question to consider is the asymptotic past behavior of
the cosmological solutions under study. As we have already mentioned, field equations
are symmetric under time reversal, and the asymptotic past corresponds to taking 
$t\rightarrow\infty$ in eqs. \family\ and \dilaton. Recently, Buonanno {\it et al.} 
\refs\rbguv\ have suggested
that the Milne Universe could represent a past attractor for the cosmological models 
with negative
spatial curvature, provided departures from homogeneity and isotropy are
sufficiently small. In our view, it is important to stress that the key ingredient in
the whole argument must be the smallness and special character of these deviations, 
for one cannot really expect
a generic cosmological model to approach Milne geometry in the asymptotic past in the
presence of the massless fields typical of string theory (see as well \refs\rn). In fact, 
one may invoke the same
arguments used in ref. \refs\rch\ to discard Misner's chaotic cosmology program on 
the grounds 
that the measure of anisotropic cosmological solutions, let alone the inhomogeneous ones, 
approaching
the isotropic regime (Milne in this case) is zero. The models we have looked at with $k=-1$
fall into the class with sufficiently isotropic initial data 
in the sense of ref. \refs\rbguv, since they approach a wedge of Minkowski 
space-time in the distant past where the dilaton is constant. This is not a suprise after all, 
because these models are in fact of Bianchi type III, belonging to the class of 
cosmologies recently studied in ref. \refs\rclt. 

For the spatially flat case ($k=0$), we are dealing with a locally rotationally symmetric
Bianchi type I model and, in this case, the line element does not approach a vacuum solution in 
the infinite past, as can be seen from the fact that the dilaton blows up in that limit.

The case $k=1$ represents quite a different physical situation. Though these models seem rather
unsuitable to describe a Pre-big Bang universe due to the fact that they 
collapse at a finite time in the past, yet the solution might be of particular interest 
in the study 
of gravitational collapse. In the special case when $\alpha=\beta=k=1$ the metric \family\
is just the Schwarschild solution inside the horizon
\eqn\sch{
ds^2=-{dt^2 \over \left({2\eta\over t}-1\right)} + 
\left({2\eta\over t}-1\right) dx^2 + 
t^2(d\theta^{2}+\sin^2\theta d\varphi^{2})
}
in the presence of a constant dilaton field. One may wonder whether it is possible 
to apply a kind of
Pre-big Bang ``graceful exit'' solution in order to regularize within string theory the
semiclassical black hole singularity at $t=0$ ($r=0$ in Schwarschild coordinates).
Then, the solution across the singularity would be related by scale factor duality
and time reversal to the solution \sch\ an will be given by ($t<0$)
$$
ds^2={-dt^2+dx^2 \over \left({2\eta\over -t}-1\right)} + 
t^2(d\theta^{2}+\sin^2\theta d\varphi^{2})
$$
along with a non-trivial dilaton field
$$
\phi=-{1\over 2}\log\left({2\eta\over -t}-1\right).
$$
A potential problem with this picture is that the dilaton goes to minus infinity as
$t\rightarrow 0^{-}$, thus resulting in a weakly coupled string theory near the singularity. 
This makes it difficult, though not impossible, to invoke non-perturbative string physics
as the smoothening mechanism near $t=0$. The problem might
be bypassed, in principle, for spherically symmetric black holes in more than six 
space-time dimensions, where
a $SL(2,{\bf R})$ transformation introducing the axion may invert the string coupling. 
At any rate, the presence of a 
curvature singularity implies that the perturbative expansion in powers of $\alpha'$
breaks down and some kind of non-perturbative worldsheet physics has to be incorporated.

To summarize, we have presented an exact and simply parametrized class 
of Kantowski-Sachs cosmological models
in dilaton gravity and studied them in the context of the Pre-big Bang scenario. We have
found that there is a narrow range of parameters leading to Pre-big Bang inflation and
that the introduction of a particular class of Type-IIB axions via a $SL(2,{\bf R})$ 
rotation may reduce the range of parameters compatible with inflation to an empty set. 
Generically, though, it seems that inflation will survive the presence of the Type-IIB axion.
We have also briefly discussed the asymptotic past behavior of these models, and 
have concluded that the 
open models ($k=-1$) have a Milne geometry as a past attractor. Contemplating
with the posibility of extending the ideas of Pre-big Bang scenario to a closed case
($k=1$), we were led to speculate about the fate of space-time beyond the Schwarzschild 
singularity. In this line of thought, provided there exists a graceful exit solution to 
overpass the Big Bang singularity there is no reason to discard a similar mechanism
to smoothly extend space-time beyond the singularities arising in gravitational collapse.

\newsec{Acknowledgments}

We are indebted to I.L. Egusquiza, R. Lazkoz, J.E. Lidsey, J.L. Ma\~nes and R. Tavakol for 
useful communications.  A.F. was supported by a Spanish Science Ministry Grant 
172.310-0250/96 and M.A.V.-M. by a University of the Basque Country Grant 
UPV-EHU-063.310-EB225/95 and a Basque Goverment Postdoctoral Fellowship.

\listrefs

\bye